\documentclass[11pt]{article}
\usepackage{epsfig}
\usepackage{amsfonts}

\def\laq{~\raise 0.4ex\hbox{$<$}\kern -0.8em\lower 0.62
ex\hbox{$\sim$}~}
\def\gaq{~\raise 0.4ex\hbox{$>$}\kern -0.7em\lower 0.62
ex\hbox{$\sim$}~}

\def\beq{\begin{equation}}
\def\eeq{\end{equation}}
\def\bea{\begin{eqnarray}}
\def\eea{\end{eqnarray}}

\def \la {\lambda}
\def \La {\Lambda}

\def \r {\rho}

\def \noi {\noindent} 

\def \Mp {M_{\rm P}}
\def \Ms {M_{\rm SUSY}}
\def \Mbu {M_{\rm S}^{\rm bulk}}
\def \d {{\rm d}}

\begin{document}
\begin{titlepage}

\begin{flushright}
BA-TH/590-08\\
arXiv:08mm.nnnn
\end{flushright}

\vspace{1.2 cm}

\begin{center}

\huge{Why supersymmetry should be restored \\ at the TeV scale}

\vspace{1cm}

\large{M. Gasperini}

\bigskip
\normalsize

{\sl Dipartimento di Fisica,
Universit\`a di Bari, \\
Via G. Amendola 173, 70126 Bari, Italy\\
and\\
Istituto Nazionale di Fisica Nucleare, Sezione di Bari, Bari, Italy \\
\vspace{0.3cm}
E-mail: {\tt gasperini@ba.infn.it}

\vspace{0.3cm}
Date: {\rm 19 March 2008}}

\vspace{1cm}

\begin{abstract}
\noi
It is explained why the curvature associated to the vacuum energy density arising from SUSY breaking cannot be completely transferred to the extra spatial dimensions of a bulk space-time manifold, and it is shown -- without using hierarchy arguments but only the results of current large-scale observations -- why the Tev scale should correspond to the maximal allowed SUSY-breaking scale. 
\end{abstract}
\end{center}

\bigskip
\begin{center}
---------------------------------------------\\
\vspace {5 mm}
{\em Essay written for the {2008 
Awards for Essays on Gravitation}}\\
{\em (Gravity Research Foundation, Wellesley Hills, MA, 02481-0004,
USA)\\
and awarded with ``Honorable Mention"}
\end{center}

\end{titlepage}

\newpage
\parskip 0.2cm

The aim of this paper is to argue that we must expect a phenomenological upper bound of about one TeV on the supersymmetry breaking scale characterizing standard-model interactions, provided $i)$ the space-time in which we live is the four-dimensional section of a curved higher-dimensional bulk manifold, and $ii)$ the vacuum energy density associated to SUSY breaking is gravitating like all other forms of energy. Let us start with two observations which are hardly questionable (at least, at the present stage of understanding of fundamental physics).

The first observation is that supersymmetry is broken, in our four-dimensional Universe, at a scale $\Ms \gaq 10^2$ GeV (as the required superpartners of known particles are not observed, up to this mass/energy scale). This necessarily produces a vacuum energy density $(\Ms/\Mp)^4 \gaq 10^{-64}$, where $\Mp = (8 \pi G)^{-1/2}\sim 10^{18}$ GeV is the reduced Planck mass (as the zero-point energies of boson and fermion fields fail to cancel one another, in general, in the absence of supersymmetry constraints).

The second observation is that such a huge energy density, in spite of its presence, {\em does not bend} the geometry of the four-dimensional macroscopic space-time that we are presently observing (as the currently observed level of curvature, at large scales, corresponds to a much smaller vacuum energy density, $\r_V/\Mp^4 \laq 10^{-120}$). Barring the (exotic) possibility that the vacuum energy associated to a SUSY-broken phase is gravitationally neutral, this second observation provides a compelling motivation for the existence of extra spatial dimensions able to absorb the curvature produced by the four-dimensional energy density $\Ms^4$, and thus   characterized by a curvature scale $L$ such that 
\beq
L^{-2} \sim 8 \pi G\, \Ms^4 ={\Ms^4\over \Mp^2}.
\label{1}
\eeq
In that case, our four-dimensional space-time may be seen as a (nearly flat) section of a (highly curved) higher-dimensional bulk manifold -- much in the same way as ordinary three-dimensional space, in spite of the energy density it contains, may correspond to a flat section of a curved FRW space-time. 

This idea of ``off-loading" the gravitational effects of the vacuum energy density along extra spatial dimensions is rather old \cite{1}, but  has been recently considered with renewed interest \cite{2}-\cite{5} within the so-called ``brane-world" scenario, where standard-model (gauge) interactions are confined on the four-dimensional hypersurface swept by the time-evolution of some fundamental three-brane. Putting aside the problems of naturalness, fine-tuning, self-tuning, \dots \,possibly associated to the existing examples of this off-loading mechanism, the crucial point (for the purpose of this paper) concerns the supersymmetry properties that the bulk must satisfy (to a high degree of accuracy) for a successful (i.e. realistic) scenario. 

In fact, if bulk supersymmetry is broken at a scale $\Mbu$, it can be shown that our four-dimensional space-time (henceforth brane-world, for short) automatically absorbs from the bulk a vacuum energy density $\r_V$, which is determined in general by $\Mbu$ and by the bulk curvature scale $L$. Such a contribution arises from the zero-point energies necessarily associated to the quantum fluctuations of the bulk geometry at the brane position \cite{6}, which are perceived on the brane as towers of massive gravitational excitations not included into the standard (four-dimensional) field multiplets. 

In order to evaluate the energy  density transferred in this way from the bulk to the brane we may consider an unperturbed background configuration describing a three-brane embedded in a $D=(4+n)$-dimensional bulk manifold, whose curvature along the $n$ extra dimensions completely absorbs the huge contribution of brane SUSY breaking, according to Eq. (\ref{1}). We stress that such a contribution arises from the quantum zero-point energies of all fields localized on the brane, possibly including the massless components of metric fluctuations (representing long-range gravitational interactions) if they are confined on the brane (for instance, through an infinitely attractive potential as in \cite{2}). 

The complete spectrum of metric fluctuations also includes, however, a wide sector of tensor/scalar perturbations which can freely propagate 
along all bulk directions, and whose oscillations evaluated at the brane position can in principle drag up energy from the bulk to the brane. The effective action for such fluctuations, dimensionally reduced (integrating over the directions orthogonal to the brane), canonically normalized, and evaluated at the brane position, describes massive tensor/scalar fields which, once quantized, generate a zero-point energy density \cite{6}:
\beq
\r_V \sim L^n \int_0^\la \d^n \chi \int_0^\La \d^3 k \sqrt{\left|k\right|^2 + 
\left|\chi\right|^2}.
\label{2}
\eeq
Here $\chi_i$, $i=1, \dots, n$ are the momenta along the $n$ extra dimensions, appearing on the brane as a (continuous) spectrum of mass eigenvalues, and  $\La$, $\la$ are the (possibly different) cutoff parameters in the momentum spaces corresponding to the spatial dimension internal and external to the brane, respectively. We have supposed, for simplicity, that the $n$ extra dimensions are isotropic (all with the same curvature scale $\sim L$), and non-compact (but ``warped" by the bulk curvature), so as to be associated to a continuous spectrum of momenta $\chi^2$. 

We can now observe that the vacuum energy (\ref{2}) of the (bosonic) metric fluctuations, in the case of a perfectly supersymmetric bulk, will be exactly compensated by the opposite contribution of the existing bulk fermionic ``superpartners". If, on the contrary, bulk supersymmetry is broken at a scale $\Mbu$, then even subtracting the fermionic contribution the net result for the vacuum energy density is in general non-vanishing. Considering, in particular, a model of broken supersymmetry with the same number of boson and fermion degrees of freedom, subtracting from Eq. (\ref{2}) the associated contribution of the fermionic components of the gravitational supermultiplet, and assuming  the so-called ``supertrace" cancellation of the mass-squared terms \cite{7} (as in models of spontaneously broken supersymmetry), we may expect from (\ref{2}) the leading order result \cite{6}
\beq
\r_V \sim \left(L \,\Mbu\right)^n \left(\Mbu\right)^4.
\label{3}
\eeq
We have assumed that $\la \sim \Mbu$, since above that scale supersymmetry is restored, and appropriate cancellations are expected to suppress to zero the vacuum energy density sourced by the bulk fields. 

It is important to stress that the above result for the vacuum energy density absorbed from the bulk by the brane is valid for 
$L\Mbu \gaq1$, while it reduces to $\r_V \sim (\Mbu)^4$ when $L\Mbu \laq1$. In other words, the result for $\r_V$ is {\em bounded from below} by 
\beq
\r_V \gaq \left( \Mbu\right)^4,
\label{4}
\eeq
quite irrespectively of the continuous or discrete nature of the spectrum of the metric fluctuations (namely, for both compact and non-compact extra dimensions). 

In fact, let us first observe that the result (\ref{3}), obtained in the case of a continuous spectrum, can also be extended to the case of compact extra dimensions (of proper size $L$) and discrete spectrum,  provided the mass-eigenvalue spacing $L^{-1}$ is negligible with respect to  the energy scale $\Mbu$ (i.e. for $L\Mbu >1$). In the opposite case ($L\Mbu <1$) in which the discrete nature of the spectrum cannot be ignored, instead, the integral $L^n \int \d^n \chi$ of Eq. (\ref{2}) is replaced by the (dimensionless) sum operator $\sum_n$ over the eigenvalues $\chi_n$, and result for $\r_V$ -- assuming the convergence of the series of SUSY-breaking corrections at the bulk scale, as before -- becomes $\r_V \sim (\Mbu)^4$. This result can also be understood as a consequence of the Casimir effect which, for $L^{-1} > \Mbu$, dominates the vacuum energy density of the compact extra dimensions, leading to an overall bulk energy density $\r_{\rm bulk} \sim (L^{-1})^n (\Mbu)^4$. The energy absorbed by the brane, given by the integration of $\r_{\rm bulk}$ over the proper volume of the transverse dimensions, is then $\r_V \sim (\Mbu)^4$. 

Finally, let us come back to the case of a continuous spectrum of bulk fluctuations, associated to warped, non-compact dimensions with a  curvature scale of order $L$, and leading to the result of Eq. (\ref{3}). If $L\Mbu >1$ such a result is always valid, quite independently of the absolute magnitude of the two energy scales $L^{-1}$ and $\Mbu$. If $L\Mbu <1$ the result (\ref{3}) is still valid, in principle, but we must take into account that a curvature $L^{-1}$ necessarily induces a bulk supersymmetry breaking characterized by a scale of the same order \cite{8,9}. Hence, in this case, bulk supersymmetry is automatically broken up to a scale $\Mbu \sim L^{-1}$, and we are led again to 
$\r_V \sim L^{-4} \sim (\Mbu)^4$.

There are two main conclusions that we can draw from the above discussion. First, when bulk supersymmetry is broken at a scale $\Mbu$, the quantum zero-point energies of the bulk gravitational fluctuations necessarily induce on the brane a vacuum energy density $\r_V$ which is {\em at least} of order $(\Mbu)^4$. Second, even in the absence of specific sources of bulk SUSY breaking, a curved geometry breaks bulk supersymmetry and thus induces on the brane a vacuum energy density $\r_V \sim L^{-4}$. On the other hand, the bulk {\em has} to be curved (according to Eq. (\ref{1})) in order to absorb the huge cosmological constant $\Ms^4$ associated to supersymmetry breaking on the brane. It follows that the process of ``off-loading" to the bulk the effects of SUSY breaking cannot have a hundred per cent efficiency: it is necessarily associated to a gravitational ``backreaction" assigning to the brane a minimal level of vacuum energy density which, according to Eq. (\ref{1}), is given (in Planck units) by
\beq
{\r_V\over \Mp^4} \sim \left(L \,\Mp\right)^{-4} \sim \left(\Ms \over \Mp \right)^8.
\label{5}
\eeq
Similar results, previously obtained with different arguments \cite{7,10,11,12}, can thus find a confirmation which is fully independent of  the number and/or compactness of the extra dimensions. 

The above result bring us to the final point of this paper. The vacuum energy density that we are presently observing in our Universe is constrained by large-scale measurements (see e.g \cite{13}) to be $\r_V/\Mp^4 \laq 10^{-120}$. If our four-dimensional macroscopic Universe is a brane-world embedded in a higher-dimensional bulk, then the SUSY breaking of the (standard-model) fields confined in four-dimensions should induce on the brane the minimal level of vacuum energy density (\ref{5}), and the phenomenological constraints on $\r_V$ should consequently determine the following upper bound on the allowed 
SUSY-breaking scale:
\beq
\Ms \laq 10^{-15} \Mp \sim 1\, {\rm TeV}.
\label{6}
\eeq

In such a case, supersymmetry effects could become directly observable already within the forthcoming collider experiments (see e.g. \cite{14}), planned to be sensitive to the TeV scale. The non-observation of supersymmetric effects at the TeV scale should  be interpreted as experimental evidence against the existence of supersymmetry, and/or against the physical relevance of the scenario of ``bulk-diluted" vacuum energy density discussed in this paper. In that case, one should possibly apply to other mechanisms for the cancellation of the vacuum energy density (see e.g. \cite{15}).

\section*{Acknowledgements}
It is a pleasure to thank Gabriele Veneziano for many useful  discussions. 


\end{document}